\documentclass[twocolumn, showpacs, preprint, amsmath, amssymb, 10pt]{revtex4-2} 

\usepackage{graphicx} 
\usepackage{dcolumn}  
\usepackage{bm}       
\usepackage{hyperref} 

\begin{document}

\title{Levitated Milligram-scale Ferromagnetic Magnetometer at Room Temperature}

\author{Yuanji Sheng}
\affiliation{National Laboratory of Solid State Microstructures and Department of Physics, \href{https://ror.org/01rxvg760}{Nanjing University}, Nanjing 210093, China}
\author{Yingchun Leng}
\affiliation{Institute of Chemical Materials, China Academy of Engineering Physics, Mianyang 621900, China}
\author{Kenan Tian}
\affiliation{National Laboratory of Solid State Microstructures and Department of Physics, \href{https://ror.org/01rxvg760}{Nanjing University}, Nanjing 210093, China}

\author{Rui Li}
\affiliation{National Laboratory of Solid State Microstructures and Department of Physics, \href{https://ror.org/01rxvg760}{Nanjing University}, Nanjing 210093, China}

\author{Yiming Chen}
\affiliation{National Laboratory of Solid State Microstructures and Department of Physics, \href{https://ror.org/01rxvg760}{Nanjing University}, Nanjing 210093, China}

\author{Dingjiang Long}
\affiliation{National Laboratory of Solid State Microstructures and Department of Physics, \href{https://ror.org/01rxvg760}{Nanjing University}, Nanjing 210093, China}

\author{Siwen Chen}
\affiliation{National Laboratory of Solid State Microstructures and Department of Physics, \href{https://ror.org/01rxvg760}{Nanjing University}, Nanjing 210093, China}
\author{Xuan He}
\email{xuan.hellen@caep.cn}
\affiliation{Institute of Chemical Materials, China Academy of Engineering Physics, Mianyang 621900, China}
\author{Peiran Yin}
\email{ypr@nju.edu.cn}
\affiliation{National Laboratory of Solid State Microstructures and Department of Physics, \href{https://ror.org/01rxvg760}{Nanjing University}, Nanjing 210093, China}


\begin{abstract}
Levitated mechanical oscillators are emerging ultrasensitive sensors with tremendous potential in both applied and fundamental physics. Levitated ferromagnets, with internal spin noises rapidly averaged, promise ultrahigh magnetic sensitivity. Here, we demonstrate a milligram-scale diamagnetically levitated ferromagnet system operating at room temperature. Through optimized geometry and multi-channel dissipation control, we achieve a magnetic sensitivity of 23~fT$/\sqrt{\text{Hz}}$ at frequency of 100-Hz level. We anticipate that a ferromagnetic magnetometer with subfemtotesla sensitivity is within reach, after modest technical improvements. This platform establishes a high-performance magnetometer for biomagnetic field detection and beyond-standard-model force searches.
\end{abstract}

\maketitle


\section{\label{sec:level1} INTRODUCTION}
High-precision magnetic sensing enables broad applications ranging from biomagnetic field detection~\cite{Glover2002} and nanoscale magnetic phenomena characterization~\cite{Cheong2020} to fundamental physics investigations such as searching for axions, dark photons, exotic spin-dependent interactions~\cite{Antypas2022,Cong2024}. While state-of-the-art magnetometers—including optically pumped atomic sensors, SQUIDs, and solid-state spin systems—now achieve subfemtotesla sensitivity~\cite{Budker2007,Schmelz2016,Zhao2023}, further advances demand new approaches. Ferromagnets, with high density of correlated electron spins locked along an easy axis via exchange interactions, offer distinct advantages on magnetic field sensitivity.

 For an underdamped system, such as a ferromagnet levitated in vacuum, the ferromagnet display unique rotational dynamics governed by their spin ensemble. When polarized electron spins dominate the angular momentum, the ferromagnet undergoes Larmor precession about external magnetic fields—behaving as a single macrospin. Theoretical work predicts such precessing ferromagnets could surpass the standard quantum limit for magnetometry employing independent spin ensembles (e.g., optical atomic magnetometer)~\cite{JacksonKimball2016,Band2018,Ni2025}. Conversely, in the more common high-angular-momentum regime, ferromagnets tend to align with the field direction, and exhibit librations about the field axis.

Levitated mechanical oscillators are emerging platforms with tremendous potential across fundamental and applied physics~\cite{GonzalezBallestero2021}. Levitated by magnetic ~\cite{Simon2001,Slezak2018,Leng2021,Fuwa2023,Tian2024}, optical~\cite{Gieseler2013,Ahn2020,Streltsov2021,Aggarwal2022}, electrical~\cite{Millen2015,Pontin2020,Dania2024,Melo2024}, and superconducting~\cite{Timberlake2019,Wang2019,Jiang2020,Hofer2023,Fuchs2024,Ahrens2025} methods, these systems achieve unprecedented environmental isolation. This unique property enables their development as ultrasensitive force sensors featuring ultralow mechanical dissipation, with broad metrological applications including gravity~\cite{Matsumoto2019,Westphal2021,Fuchs2024}, inertia\cite{Monteiro2017,Timberlake2019,Xiong2021}, torque~\cite{Ahn2020}, and electromagnetic field measurements~\cite{Frimmer2017,Gieseler2020,Delord2020,Ahrens2025}. Notably, milligram and sub-milligram diamagnetically levitated oscillators have recently demonstrated particular promise for probing beyond-standard-model forces~\cite{Xiong2021,Yin2022,Yin2025,Tian2025}. Further refinement could establish levitation platforms as novel probes of low-energy dark sectors~\cite{Carney2021,Moore2021}. Concurrently, these systems provide versatile testbeds for quantum measurement and quantum state preparation~\cite{Rossi2018,Delic2020,Magrini2021,Stickler2021,Rusconi2022,Huang2024,Jin2024}, potentially advancing studies of quantum decoherence~\cite{Kaltenbaek2023} and gravitational interactions in quantum systems~\cite{Bose2025}.

 Through intrinsic spin-projection noise suppression, the levitated ferromagnet is proposed to realize a ultrahigh magnetic sensitivity~\cite{Vinante2021}, making it a promising platform that has attracted significant research interest. Achieving high-precision levitated ferromagnetic magnetometer requires minimizing extrinsic noises and dissipations. Current development critically depends on addressing three key challenges: (i) minimization of mechanical dissipation, (ii) suppression of multi-source noise (arising from the levitation system, measurement protocols, and environment), and (iii) ensuring long-term operational stability. Recent work on Meissner-levitated ferromagnets has achieved a magnetic sensitivity of 20 fT/$\sqrt{\text{Hz}}$~\cite{Ahrens2025}, such systems require cryogenic environments, significantly increasing complexity.

In this Letter, we demonstrate a milligram-scale diamagnetically levitated ferromagnet system operating at room temperature. Through optimized geometry design and multi-channel dissipation suppression, the system achieves a magnetic sensitivity of 23~fT$/\sqrt{\rm Hz}$ near the librational resonant frequency of $\omega_0/2\pi=153.001$~Hz. Current magnetic sensitivity, limited by the vibration noises, was calibrated through dual methods: direct optical rotation angle measurement and applied periodic magnetic field excitation, with excellent mutual agreement. This sensitivity matches the state-of-the-art of cryogenic Meissner-levitated systems~\cite{Ahrens2025}. Targeted technical developments—including further suppression of vibration noise and magnetic hysteresis loss—will enable magnetic sensitivity of sub-fT in a near future. The platform's inherent cryogenic compatibility further promises significant future enhancements. These capabilities establish it as both a high-performance room-temperature magnetometer and a versatile platform for spin-mechanics coupling and quantum rotational studies.

\section{EXPERIMENTAL SYSTEM}
\subsection{Design Principle}
We achieve a stable levitation of a hard ferromagnet with a lifting magnet and a pair of diamagnetic bismuth (Bi) plate~\cite{Simon2001,Leng2024}. For a macroscopic ferromagnet levitated in a magnetic field $\bm{B}$, the gyroscopic dynamics are negligible and instead the ferromagnet will librate about the field axis. The torque $\bm{\tau}$ applied by the magnetic field is $\bm{\tau} = \bm{\mu} \times \bm{B}$. Here, $\bm{\mu} = \bm{M} V$ is the magnetic moment, with $\bm{M}$ denoting the magnetization and $V$ the volume of the ferromagnet. The resonance frequency of the librational mode of the levitated ferromagnetic mechanical oscillator is
\begin{equation}
    \omega_{\text{0}} = \sqrt{\frac{MBV}{I}},
\end{equation}
 where $I$ is the moment of inertia about its rotation axis.

The dynamic equation of the librational motion of the levitated ferromagnet is expressed as:
\begin{equation}
I\ddot{\theta}+I\gamma\dot{\theta}+I\omega_{0}^2\theta=\tau^{\rm signal}+\tau^{\rm noise}.
\label{motion}
\end{equation}
Here, $\theta$ is the angular displacement of the mechanical oscillator from its equilibrium position, $\tau^{\rm noise}$ represents the external torque noises applied to the oscillator and $\tau^{\rm signal}$ corresponds to the target signal torque. The relationship between the torque acting on the oscillator $S_{\tau\tau}(\omega)$ and the angular displacement $S_{\theta\theta} (\omega)$ is $S_{\tau\tau}(\omega)=S_{\theta\theta}(\omega)I^2/| \chi(\omega) |^2$, with $\chi(\omega)=1/(\omega_0^2-\omega^2+i \omega \gamma)$ being the mechanical response function of the mechanical oscillator. 

To detect a magnetic field perpendicular to the magnetization axis, with the signal torque being $\tau^{\rm signal} = MVB$, we get the expression of power spectral density (PSD) of the magnetic field:
\begin{equation}
S_{BB}(\omega) = \frac{S_{\theta\theta}(\omega) I^2}{M^2 V^2| \chi(\omega) |^2}.
\end{equation}
In the thermal noise limit, the PSD of torque noises follows $S_{\tau\tau}^{\text{th}} = 4\gamma k_{\text{B}} T I$, according to the fluctuation-dissipation theorem \cite{kubo1966fluctuation}, where $k_{\text{B}}$ is Boltzmann's constant, and $T$ is the temperature. The corresponding thermal-noise-limited magnetic sensitivity is 
\begin{equation}
\sqrt{S_{BB}^{\text{th}}} = \sqrt{\frac{4\gamma k_{\text{B}} T I}{M^2 V^2}}.
\label{eq:sensitivity}
\end{equation}
Key sensitivity enhancement strategies therefore target: (i) $\gamma$ suppression, (ii) $I/V^2$ minimization via geometric optimization, and (iii) external noise mitigation. Cryogenic environment further leverages the explicit $T$-dependence in Eq. \eqref{eq:sensitivity}.

\subsection{Experimental Setup}
\begin{figure*}
    \centering
    \includegraphics[width=0.9\textwidth]{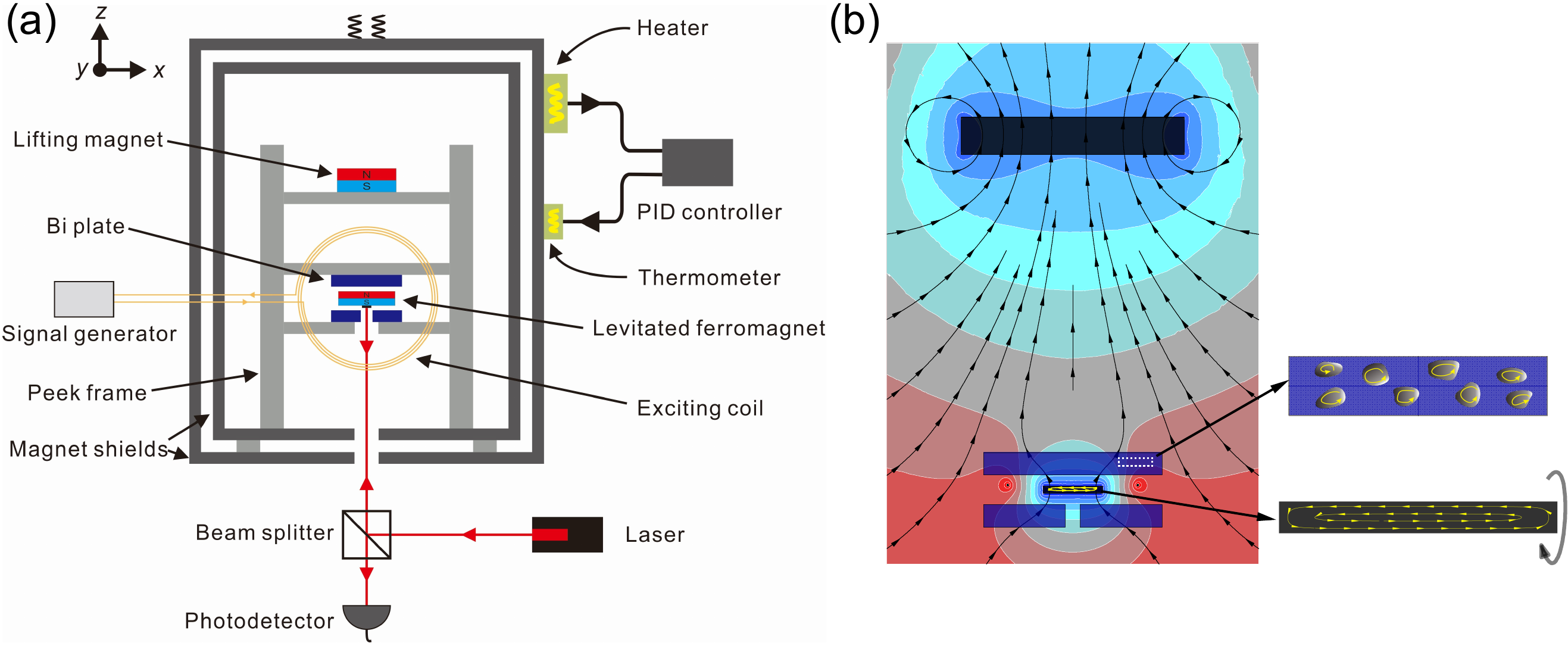}
    \caption{
    (Color online) Schematic of the diamagnetically levitated ferromagnet system. (a) An elongated rectangular hard ferromagnet is levitated by a lifting magnet and a pair of Bi plates, placed on a PEEK frame. The Bi plates made of bismuth nanoparticles and cetyl alcohol serves as a diamagnetic unit, providing a stable potential well for the levitated ferromagnet. Two layers of ferrite magnetic shield outside the PEEK frame are used to shield external magnetic field noises and suppress eddy currents generated in metallic conductors. The thermometer and heater fixed on the ferrite magnetic shields are used to stabilize the environment temperature via a PID controller. A pair of excitation coils are positioned in front of and behind the levitated ferromagnet for magnetic sensitivity calibration. (b) Simulations of the magnetic field distribution in the diamagnetic levitation system and the corresponding eddy current distributions (yellow curves) in Bi plates and the levitated ferromagnet. The naturally formed oxide layers on Bi nanoparticles' surfaces effectively confine eddy currents to the regions within the particles, significantly suppressing the eddy current dissipation. 
    }
    \label{fig:1}
\end{figure*}

The experimental setup is depicted in Fig.~\ref{fig:1}(a). The levitating magnet is a permanent magnet made of neodymium-iron-boron (NdFeB) material, chosen for its high magnetization. To minimize $I/V^2$, the ferromagnet is designed as an elongated bar with dimensions of $1\,\text{mm} \times 1\,\text{mm} \times 8\,\text{mm}$, with mass $60.4 \pm 0.1$~milligram. The longitudinal axis of the ferromagnet is aligned with the rotational axis ($x$-axis), while its magnetization is oriented vertically upward along the $z$-axis. A $100\,\mu\text{m}$-thick silicon strip ($1\,\text{mm} \times 7.7\,\text{mm}$) is UV-bonded beneath the ferromagnet for optical detection.  

The lifting magnet with a radius of 15\,mm and thickness of 5\,mm is primarily used to counteract the gravity of the magnetic oscillator. Two diamagnetic bismuth plates (\( 17.5\,\text{mm} \) radius, \( 3\,\text{mm} \) thickness) consist of metal Bi nanoparticles (typical particle size: ~50\,nm) and cetyl alcohol are positioned above and below the levitated ferromagnet, respectively. The lower Bi plate features a concentric aperture (\( 1.5\,\text{mm} \) radius) for optical access. The potential well formed by the interplay of gravitational, magnetic (from the lifting magnet), and diamagnetic (from Bi plates) potentials enables stable levitation of the magnetic oscillator. 

Conventional diamagnetic levitation systems typically employ pyrolytic graphite as the diamagnetic material. While graphite provides stable trapping potential, it also generates significant eddy currents that become the dominant dissipation source. 
In this experiment, bismuth serves as the diamagnetic material due to its strong diamagnetic response and spontaneous formation of insulating oxide layers upon air exposure. As shown in Fig.~\ref{fig:1}(b), this native oxide layer effectively blocks inter-particle electron transfer, significantly suppressing eddy current generation. We fabricated high-resistivity diamagnetic plates by mixing Bi nanoparticles with molten cetyl alcohol followed by solidification. The Bi-based system demonstrates significant potential for minimizing eddy current dissipation.

The entire levitation system is housed within a frame made of polyether ether ketone (PEEK), a kind of insulating material, to avoid the formation of eddy current. Two exciting coils (inner radius: 15.0\,mm, outer radius: 18.5\,mm, 100 turns each) are mounted on the front and rear sides of the PEEK frame with a 50\,mm axial spacing, forming a calibration system where the magnetic oscillator is precisely centered between the coils. A two-layer soft-magnetic MnZn ferrite shield is used to shield magnetic noise and suppress eddy currents in conductors outside the shield, owing to its high permeability and electrical resistivity. The thermometer and heater are installed outside the magnetic shield. We stabilize the temperature of the levitated ferromagnet with a proportional integral derivative (PID) controller, to long-term operational stability of our system.  In addition, the entire setup is placed on a three-stage spring-mass suspension system to isolate external vibrations.

A 532-nm laser beam is incident vertically onto the silicon stripe beneath the levitated ferromagnet. After passing through a 0.38~m long optical lever, the reflected light is then detected by a four-quadrant photodetector, which outputs differential analog voltage signals \( V_{\text{diff},X} \) and \( V_{\text{diff},Y} \), along with a sum signal \( V_{\text{sum}} \). 
By adjusting the detector's position, the rotational angles $\theta_x$ and $\theta_y$ of the levitated ferromagnet about its longitudinal ($x$-) and transverse ($y$-) axes can be characterized by the differential voltage signals $V_{\mathrm{diff},Y}$ and $V_{\mathrm{diff},X}$, respectively (see Appendix~\ref{Optical Detection} for more details).

\section{Characters of levitated ferromagnets}

\begin{figure}[h]
    \hspace*{-1cm} 
    \includegraphics[width=0.85\linewidth]{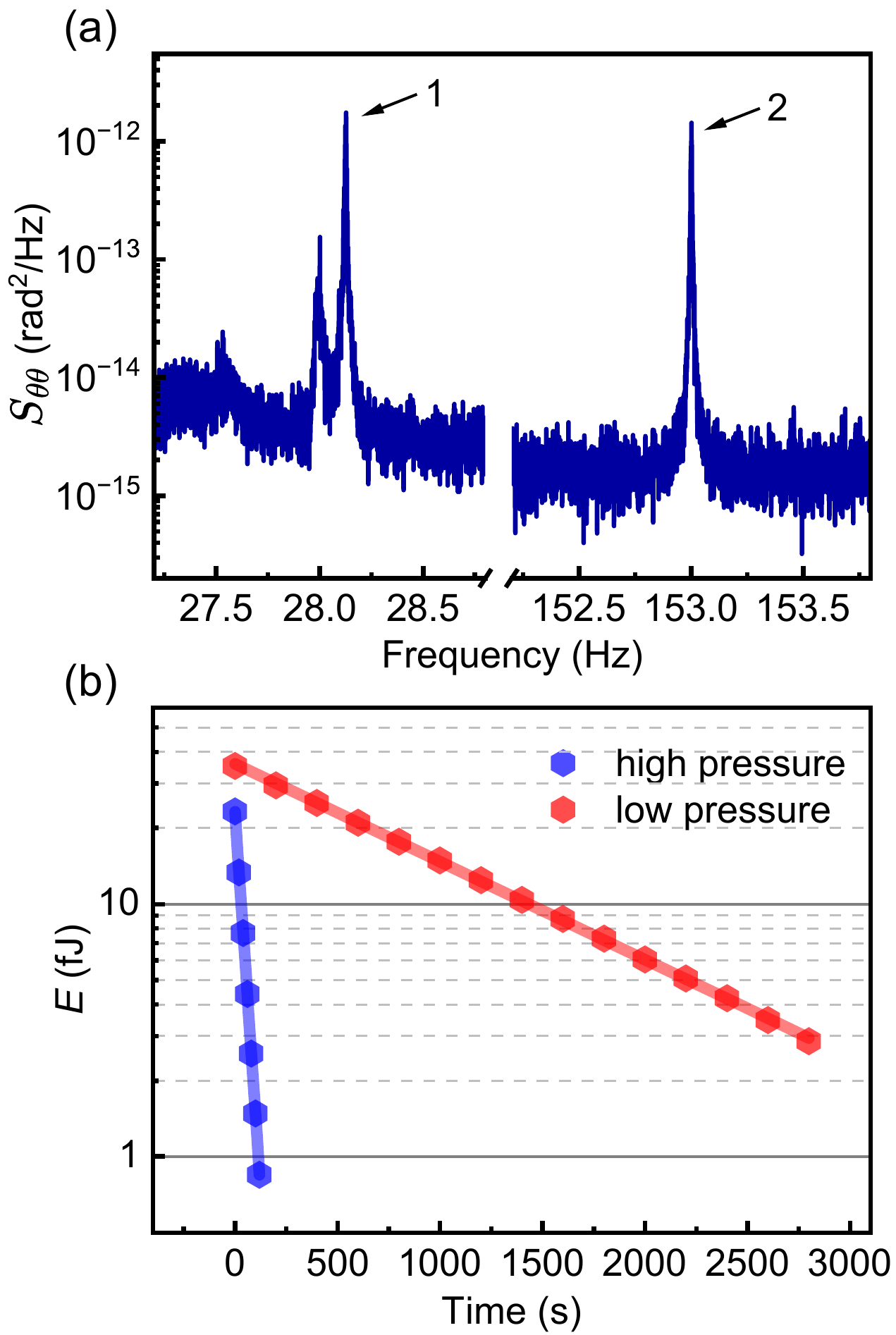}
    \caption{The frequency and mechanical dissipation of the levitated ferromagnets. (a) The PSD of levitated ferromagnet's rotation angle. A continuous measurement of total time 7200~s is divided into 6 equal segments to derive the averaged PSD. Peak~1 and Peak~2 correspond to the rotational modes around the $y$-axis and $x$-axis, with resonant frequencies of 28.126\,Hz and 153.001\,Hz, respectively. (b) The ringdown measurements of the $x$-axis rotational mode. Blue and red dots depict the oscillator's energy decay under high and low pressure, respectively. Blue and red lines show the exponential decay fitted curve, revealing decay times of 36.2 s and 324.3 s for the high- and low-pressure regimes, respectively.}
    \label{fig:2}
\end{figure}

To characterize the rotational dynamics of the oscillator, we define the total power spectral density \( S_{\theta\theta} = S_{\theta\theta}^x + S_{\theta\theta}^y \), where \( S_{\theta\theta}^x \) and \( S_{\theta\theta}^y \) represent the power spectral densities of the rotation angles \( \theta_x \) (long-axis rotation) and \( \theta_y \) (short-axis rotation), respectively. Fig.~\ref{fig:2}(a) shows the measured \( S_{\theta\theta} \) at a pressure of \( 8.6 \times 10^{-2} \) mbar ( denoted as high pressure hereafter). The power spectral density (PSD) data were obtained by dividing a 7200-second continuous measurement into six equal segments, computing individual PSDs, and averaging. Peak~1 and Peak~2 correspond to the rotational modes around the $y$-axis (28.126\,Hz) and $x$-axis (153.001\,Hz), respectively. Three translational modes ($<$5\,Hz) are not shown in the figure. And the rotational mode around $z$-axis exhibits low sensitivity with employed optical detection scheme. In this experiment, we focus on the $x$-axis rotation mode (Peak~2), and determine the magnetic sensitivity via its angular response. Fitting the the $x$-axis rotational PSD to a Lorentzian lineshape
$\sim\frac{\gamma/2\pi}{4(f-f_0)^2 + (\gamma/2\pi)^2}$, yields a full width at half maximum (FWHM) $\gamma/2\pi= 4.40\,\text{mHz}$, corresponding to a Q factor of $3.5\times10^4$. This agrees with ringdown measurements (Fig. ~\ref{fig:2}(b), blue data).

The librational mode resonant frequency of the levitated ferromagnet is given by $\omega_{\text{0}} = \sqrt{\frac{MBV}{I}}$.
Since the levitation position of the magnetic oscillator depends on the gradient of the magnetic field B along the z-direction rather than its magnitude, we can control the oscillator’s frequency without displacing the equilibrium position by applying a uniform $z$-directional magnetic field using Helmholtz coils. Moreover, for the oscillator employed in this experiment, the ratio \( V/I \) is inversely proportional to \( a^2 + b^2 \), where \( a \) and \( b \) represent the length and width of the oscillator's cross-section perpendicular to the rotation axis. Consequently, the oscillator's resonant frequency can also be adjusted by modifying its transverse dimensions.

Figure~\ref{fig:2}(b) displays ringdown measurements of $x$-axis rotational energy decay at different pressures. At a pressure of $8.6\times 10^{-2}$ {mbar} (denoted as high pressure), the measured decay time constant is 36.2\,\text{s}, consistent with the Lorentzian-fitted mechanical dissipation $\gamma_\text{low}/2\pi$ in Fig.~\ref{fig:2}(a). At this high pressure, the mechanical dissipation are dominated by the background gas. Reducing pressure to \( 6.4 \times 10^{-5} \)~mbar (denoted as low pressure) yields a limiting decay constant of 324.3\,{s}, corresponding to $\gamma_\text{high}/2\pi=0.49\,\text{mHz},~Q= 3.1\times10^5 $. This dissipation rate is verified to be pressure-independent.

We have systematically analyzed the energy loss channels in our experimental system at low pressure, including the eddy current dissipation and magnetic hysteresis dissipation.
For a ferromagnet rotating at an angular velocity $\bm{\omega}$ in an external bias field $\bm{B}_{\text{bias}}$, the internal eddy-current-induced dissipation is given by:
\begin{equation}
\label{eq:eddy_current}
\gamma_{\text{eddy}}^{\text{fm}} = \int_{V} \frac{|\bm{J}|^2}{\sigma I \omega^2} \, \mathrm{d}V,
\end{equation}
where $\bm{J} = \sigma \bm{v} \times \bm{B}_{\text{bias}}$ is the current density, $\sigma$ is the ferromagnet's electrical conductivity, $\bm{v} = \bm{\omega} \times \bm{r}$ is the velocity of the ferromagnet's volume element $\mathrm{d}V$, and $\bm{r}$ is the position vector of the volume element $\mathrm{d}V$ in the oscillator. Finite element analysis (FEA) shows that the dissipation caused by the eddy current in the ferromagnet is $\gamma_{\text{eddy}}^{\text{fm}}/2\pi \approx 0.01~\text{mHz}$, which is significantly lower than the total dissipation $\gamma/2\pi = 0.49~\text{mHz}$. Eddy currents also appears in the surrounding conductors. We have added a two-layer ferrite magnetic shield to suppress it. For the present setup, we analyze and simulate the relative motion between the oscillator and conductors, and get the eddy current dissipation as $\gamma_{\text{eddy}}^{\text{out}}/2\pi\approx 3\times10^{-7}~\text{Hz}$. For the eddy currents in the diamagnetic Bi plate, the surface oxidation of metallic Bi particles can effectively confine these eddy currents within individual particles, significantly reducing the resulting eddy current losses. Regarding eddy current loss only within the nanoparticles, the corresponding dissipation is $\gamma_{\text{eddy}}^{\text{Bi}}/2\pi \approx 3\times10^{-10}$ Hz (see Appendix~\ref{Bi Dissipation} for details).

With eddy current dissipation efficiently suppressed, magnetic hysteresis loss emerges as the dominant dissipation mechanism~\cite{Han1995}. We analyze this loss in both the ferrite magnetic shield and the levitated ferromagnet (detailed derivation in Appendix~\ref{magnetic hysteresis loss}). Modeling the magnetic oscillator as an equivalent current-carrying coil, the magnetic hysteresis dissipation induced by the two-layer ferrite magnetic shield is $\gamma_{\text{hyst}}^{\text{shield}}/2\pi \approx 0.03~\text{mHz}$. Within the levitated ferromagnet, hysteresis losses originate from its imaginary magnetic susceptibility. For our setup, we find $\gamma_{\rm hyst}^{\rm fm} = \frac{{\mu}''B_0^2I\omega_0^3}{\mu_0^2M^2V} \propto \frac{{\mu}''(a^2+b^2) \omega_0^3}{M^2}$, where $\mu''$ is the imaginary permeability component, $a$ and $b$ the transverse dimensions. Numerical calculations give $\gamma_{\text{hyst}}^{\text{fm}}/2\pi \sim 0.1~\text{mHz}$, matching the measured total mechanical dissipation $\gamma/2\pi$. Further suppression of $\gamma_{\text{hyst}}^{\text{fm}}$ requires: (i) reducing $\mu''$, (ii) lowering $\omega_0$, (iii) minimizing transverse dimensions, and (iv) increasing $M$. Implementing these approaches would enable achieving mechanical dissipation rates approaching $\sim 10~\mu$Hz (corresponding to $Q \sim 10^7$ at 100 Hz), corresponding to a thermal-limited magnetic sensitivity of 0.3 fT/$\sqrt{\text{Hz}}$.

\section{Magnetic field detection sensitivity}
\subsection{External Noises}

In the practical measurements, the external noises mainly arise from thermal noises, environmental vibration noises, and electrical-magnetic background noises. At present, the dominate dissipation has been determined as the magnetic hysteresis dissipation in the levitated ferromagnet, the thermal noise PSD is estimated as $S_{\tau\tau}^{\text{th}} = 4\gamma k_{\text{B}} T I$. During the measurements, we have monitored the environmental vibration noises on the experiment platform with a commercial accelerator, we obtained a vibration noise  on the order of $10^{-6}\, \text{g}/\sqrt{\text{Hz}}$ around 100 Hz. In this experiment, a three-stage spring-mass suspension system has been built to isolate the external vibrations.  In addition, the electrical-magnetic background noises can be well shielded with the magnetic shields, as demonstrated in our previous experiments~\cite{Yin2022}.

\subsection{Experimental Results}
\begin{figure}[]
\centering
\includegraphics[width=0.48\textwidth]{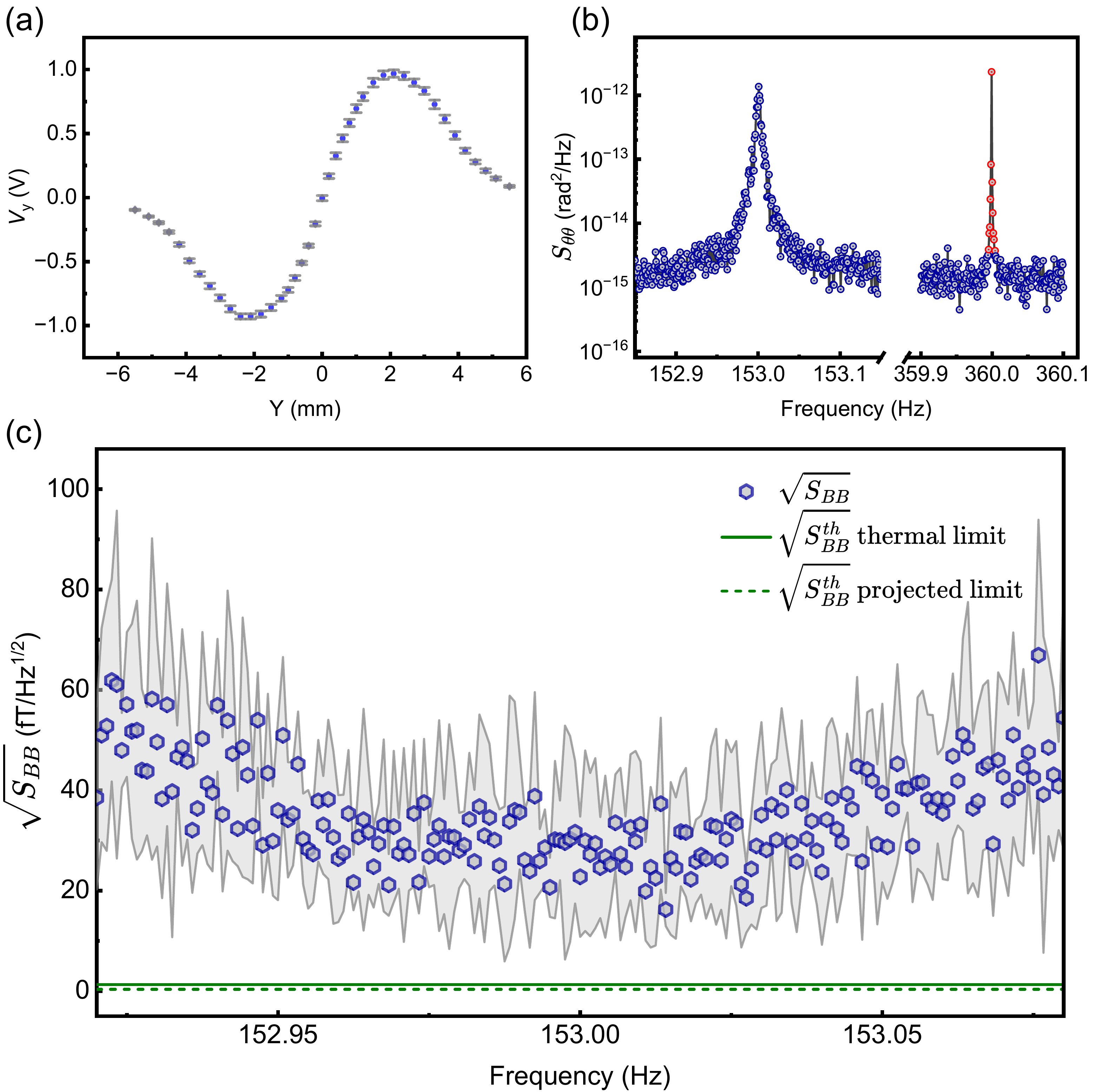}
\caption{Experiment Results. 
(a) The measured response curve of $V_{\text{diff},Y}$ to the $y$-directional displacement of the reflected light spot. The response curve has good linearity in the range of $\pm 0.2~\mathrm{mm}$($R^2=0.998$).
(b) The measured PSD of the levitated ferromagnet's angular displacement. Data points represent averages from six 1200-s measurement segments. Red points denote the response to the off-resonance magnetic drive.
(c) Magnetic sensitivity of the levitated ferromagnet determined by the measured angular displacement. The blue points represent the averaged magnetic sensitivity and gray shaded band indicates the standard deviations. The solid and dashed green curves indicate the thermal-noise-limited and projected magnetic sensitivity, respectively.
}
\label{fig:3}
\end{figure}

The signal-to-noise ratio is determined as $\mathrm{SNR}= S_{\theta\theta}^{\rm signal}(\omega)/S_{\theta\theta}^{\rm noise}(\omega)$. We adopt $\mathrm{SNR}=1$ as the criterion for distinguishing the signals from noises. The magnetic sensitivity is defined as the square root of the measured PSD of noises, expressed as $\sqrt{S_{BB}(\omega)} = \frac{I\sqrt{S_{\theta\theta}^{\rm noise}(\omega)} }{\mu | \chi(\omega) |}.$

In this experiment, the PSD of angular displacement $S_{\theta\theta}$ was determined by measuring the position of the reflected light spot using a photodetector (see Appendix~\ref{Optical Detection} for details). The $x$-axis rotational angles is determined from the measured $V_{\text{diff},Y}$. Fig. \ref{fig:3}(a) shows the measured response curve of $V_{\text{diff},Y}$ to the displacement of the reflected light spot relative to the detector center in the y direction. The curve was obtained from five repeated measurements. Within a linear response range of $\pm 0.2~\mathrm{mm}$, we conduct the angular displacement measurements.  For the oscillator operating at resonance frequency $\omega_0/2\pi = 153.001~\mathrm{Hz}$ in this experiment, this range corresponds to an angular variation of $\pm2.6\times10^{-4}~\mathrm{rad}$, which is four orders of magnitude larger than the thermal fluctuations $\sqrt{k_{\mathrm{B}} T / I \omega_0^2} = 2.1 \times 10^{-8}~\mathrm{rad}$ at room temperature. Linear fitting yields a response coefficient $k_y = \mathrm{d}V_{\mathrm{diff},Y}/\mathrm{d}y = 937 \pm 25~\mathrm{V}/\mathrm{m}$ with $R^2 = 0.998$.

With the calibrated angular displacement, we obtain the \( S_{\theta\theta} \) of $x$-axis rotational mode at a pressure of \( 8.6 \times 10^{-2} \) mbar, as shown in Fig.~\ref{fig:3}(b). To verify the magnetic sensitivity, an ac magnetic field of 360 Hz --deliberately detuned from the oscillator's resonant frequency--was applied to the levitated ferromagnet by a pair of excitation coils during measurements. The two excitation coils are connected in series and driven by a current with an amplitude of 0.197\,$\mu\mathrm{A}$. It is noteworthy that due to the magnetization of the magnetic shield by the coil-generated magnetic field, the amplitude of calibration signal $B_{\text{cal}}$ is enhanced by a factor $\alpha_{\text{shield}} =1.3$. The corresponding numerical simulation result is $B_{\text{cal}}=0.29\pm 0.02$ nT, where the standard deviation of the magnetic field stems from positional deviations of the excitation coils. Consequently, the theoretical root-mean-square (RMS) torque on the oscillator induced by the calibration signal is $\sqrt{\langle\tau_{\mathrm{theory}}^2\rangle} = \frac{\sqrt{2}}{2}M V B_{\text{cal}} = (1.9\pm0.1)\times 10^{-12}$ $\mathrm{N \cdot m}$. On the other hand, we can determine the torque from the measured angular displacements given in Fig. ~\ref{fig:3}(b), through $\langle \tau_\text{exp}^2 \rangle=\frac{1}{2 \pi}\int S_{\theta\theta}(\omega)I^2/| \chi(\omega) |^2\mathrm{d}\omega$. By integrating the off-resonance excitation spectrum after noise floor subtraction, we obtain $\sqrt{\langle\tau_{\mathrm{exp}}^2\rangle} = (2.0\pm0.1)\times 10^{-12}$ $\mathrm{N \cdot m}$, showing excellent agreement with theoretical predictions. We thereby validate the magnetic sensitivity and confirm the operational efficacy of the levitated ferromagnet platform.

Figure~\ref{fig:3}(c) presents the result of magnetic sensitivity $\sqrt{S_{BB}(\omega)}$ determined by the measured angular displacements, where the blue dots represent the averaged data and the gray shaded bands indicates the corresponding standard deviations. Near the resonant frequency, we achieved a magnetic sensitivity of $23$~fT/$\sqrt{\text{Hz}}$, which is limited by vibration noise rather than thermal noise. Away from resonance, the magnetic sensitivity degrades due to the dominance of measurement noises. The solid green curves in Fig.~\ref{fig:3}(c) show the theoretical thermal-noise-limited magnetic sensitivity at low pressure, corresponding to $2~\mathrm{fT}/\sqrt{\mathrm{Hz}}$. And the dashed green line indicates the projected sensitivity $0.3~\mathrm{fT}/\sqrt{\mathrm{Hz}}$.

\section{DISCUSSIONS AND CONCLUSIONS}
In this work, we have constructed a system of milligram-scale levitated ferromagnet operating at room temperature, and demonstrated a magnetic sensitivity of 23~fT/$\sqrt{\text{Hz}}$ through optimal geometry design and mechanical dissipation suppression. Environmental vibration noise currently constitutes the dominant limitation on magnetic sensitivity. To address this, we propose a dual approach. On the one hand, upgrade the multi-stage isolation system to further suppress the noises. On the other hand, we anticipate that it is the the large-amplitude translational modes (frequencies below 5 Hz) perturbs the $x$-axis rotational modes. Passive isolation proves ineffective at these low frequencies. Potential solutions include, increasing translational mode frequencies through Bi plate geometry engineering and implementing feedback cooling on these modes.

Systematic analysis of dissipation channels identifies magnetic hysteresis loss in the levitated ferromagnet as the dominant dissipation mechanism. Mitigation strategies, such as targeted material engineering (selecting lower-loss magnetic materials) and structural optimization, can suppress this loss effectively. Near-term technical improvements should achieve mechanical dissipation rates of $\sim10~\mu\text{Hz}$ (corresponding to $Q \sim 10^7$ at 100 Hz). Combined with the reduction of vibration noise sources discussed previously, this projects a room-temperature magnetic sensitivity of 0.3~fT/$\sqrt{\text{Hz}}$ within reach. Furthermore, the system's inherent cryogenic compatibility paves the way for significant future sensitivity enhancements.

The levitated ferromagnet platform developed in this work offers significant potential for fundamental physics exploration. With ultrahigh magnetic sensitivity, we are able to search for the efficient magnetic field generated by the ultralight Bosons~\cite{Higgins2024,Kalia2024}, such as axions and dark photons, which are the dark matter candidates. Additionally, the system permits tests of exotic spin-dependent interactions through controlled source-sensor experiments~\cite{Su2024,Huang2024a,Tian2025}. Furthermore, it provides a pathway to probe couplings between single spin and rotations~\cite{Rusconi2017,Gieseler2020,Delord2020,Rusconi2022,Ahrens2025a}, and rotational superpositions~\cite{Stickler2021,Jin2024}.

\textit{Note added.} Recently, we became aware of similar independent work by Wei Ji et al.~\cite{Ji2025}.

\begin{acknowledgments}
This work was supported by the National Natural Science Foundation of China (Grant No. T2388102), and the Natural Science Foundation of Jiangsu Province (Grant No. BK20241255).

Y. S., Y. L. and K. T. contributed equally to this work.
\end{acknowledgments}

\section*{Data Availability}
 
The data that support the findings of this article are openly
available~\cite{raw_data}, embargo periods may apply.

\appendix
\counterwithin{figure}{section}
\section{Optical detection method}

\label{Optical Detection}

\begin{figure}
    \centering
    \includegraphics[width=0.5\linewidth]{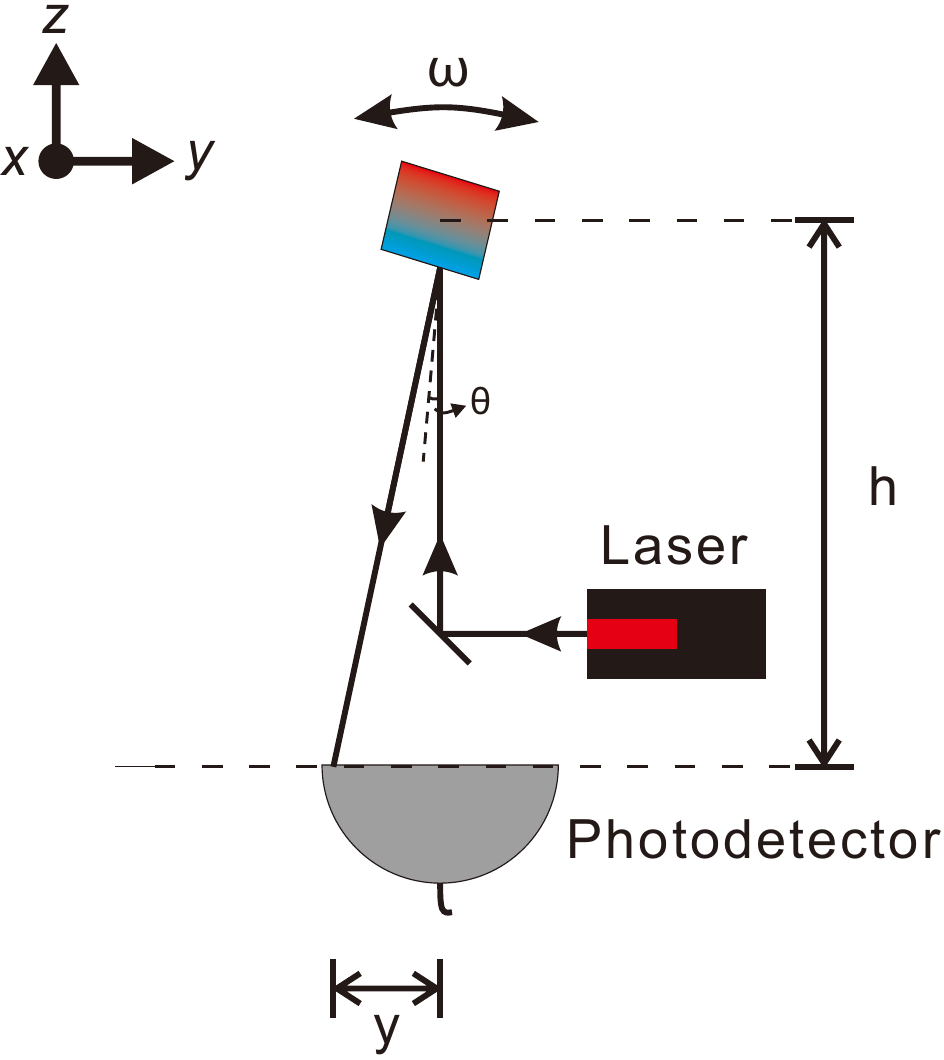}
    \caption{The scheme of optical detection in this experiment.}
    \label{fig:Optical Detection}
\end{figure} 
Figure~\ref{fig:Optical Detection} illustrates the angular displacement detection principle using a four-quadrant photodetector (4QPD). The 4QPD comprises four identical photodiodes arranged in a ${2\times2}$ matrix configuration. When operating the 4QPD in \textit{monitor mode}, the system generates three output signals: bottom-minus-top ($Y$-axis aligned) and ( left-minus-right ($X$-axis aligned) differential signals, along with a total intensity signal. The two differential signals are analog voltage quantities representing the light intensity difference sensed by paired photodiode units in the array, denoted as $V_{\text{diff},X}$ and $V_{\text{diff},Y}$. When the oscillator is in its equilibrium position, the reflected light spot is centered on the detector, resulting in $V_{\text{diff},X}^0 = V_{\text{diff},Y}^0 = 0$.
By aligning the detector such that its $X-Y$ axes coincide with the system's $x-y$ coordinate axes, the oscillator's rotation angles about its longitudinal axis ($x$-axis) and transverse axis ($y$-axis), denoted $\theta_x$ and $\theta_y$ can be characterized by:
\begin{equation}
    \theta_{x}=\frac{V_{\text{diff},Y}-V_{\text{diff},Y}^0}{(dV_{\text{diff},Y}/dy)\cdot(dy/d\theta_{x})},
\end{equation}
\begin{equation}
    \theta_{y}=\frac{V_{\text{diff},X}-V_{\text{diff},X}^0}{(dV_{\text{diff},X}/dx)\cdot(dx/d\theta_{y})},
\end{equation}
where x and y represent the position coordinates of the reflected light spot relative to the detector center. Since the distance h between the detector and levitated ferromagnet significantly exceeds the oscillator's dimensions, the angular conversion factors simplify to $dx/d\theta_{y}=dy/d\theta_{x}=2h$. 

\section{Eddy current dissipation in Bi plate}
\label{Bi Dissipation}
To analyze the eddy current dissipation in diamagnetic Bi, we consider the relative motion between the oscillator and the Bi plate and calculate the eddy current losses in the diamagnetic Bi plate when it rotates around the oscillator's axis of rotation ($x$-axis).  

Since the oxide layer on the surface of metallic Bi particles insulates them from each other, we analyze the eddy currents within a single metallic Bi nanoparticle.  
For a sphere of radius $R$ moving with velocity $\bm{v}$ in a magnetic field $\bm{B}$, the eddy current dissipation power is given by \cite{Chen2022,Xie2023} 
\begin{equation}
\label{P_sphere}
    P_{\text{sphere}} = \frac{2\pi R^5}{15 \rho_\text{r}}(\frac{\mathrm{d} B_v}{\mathrm{d} t})^{2},  
\end{equation}  
where $\rho_\text{r}$ is the electrical resistivity of the nanosphere, $B_v=\mathbf{B}\cdot \frac{\mathbf{v}}{| \mathbf{v} |}$ is the component of the magnetic field $\bm{B}$ along the direction of velocity $\bm{v}$. 

When the oscillator rotates about its long axis ($x$-axis) with angular velocity $\bm{\omega}$, the metallic Bi particles in the nearby diamagnetic Bi plate move with relative velocity $\bm{v} = -\bm{\omega} \times \bm{r}$ with respect to the oscillator. Under these conditions, the term $\mathrm{d}B_v/\mathrm{d}t$ in \eqref{P_sphere} can be expressed as
\begin{equation}
    \frac{\mathrm{d} B_v}{\mathrm{d} t} = \omega\left(\frac{\mathrm{d} f}{\mathrm{d} y}z - \frac{\mathrm{d} f}{\mathrm{d} z}y\right),
\end{equation}
where $f(x,y,z) = \frac{z B_y - y B_z}{\sqrt{y^2 + z^2}}$ is a function related to the position of the metallic sphere.

Consequently, the eddy current power dissipation per unit volume of sphere is
\begin{equation}
P_{\text{unit}} = \frac{\omega^2 R^2}{10 \rho_{\text{r}}}\left(\frac{\mathrm{d} f}{\mathrm{d} y} z - \frac{\mathrm{d} f}{\mathrm{d} z} y\right)^2.
\end{equation}

Finally, the resulting dissipation of the magnetic oscillator due to the diamagnetic Bi plate with a metal particle volume fraction of $\eta$ is
\begin{equation}
    \label{eq:dissipation_eddy}
\gamma_{\text{Bi}} = \frac{\int_V P_{\text{unit}} \eta \mathrm{d}V}{I \omega^2} = \frac{\eta R^2}{10 I \rho_{\text{r}}}\int_V \left(\frac{\mathrm{d} f}{\mathrm{d} y} z - \frac{\mathrm{d} f}{\mathrm{d} z} y\right)^2 \mathrm{d}V,
\end{equation}
where $I$ is the moment of inertia of the magnetic oscillator about its long axis ($x$-axis). It is noteworthy that the parameter $\eta$ is set to unity in the simulation for a conservative estimate.
\section{Magnetic hysteresis loss}
\label{magnetic hysteresis loss}

For a magnetic medium in an oscillating field \( \mathbf{B} \), the average dissipation power due to hysteresis effects is given by~\cite{Han1995}
\begin{equation}
    \bar{P} = \int_V \frac{1}{2} \omega \mu'' H^2 \, \mathrm{d}V,
\end{equation}
where \( \omega = 2\pi f \) is the angular driving frequency, \( \mu'' \) is the imaginary part of the complex permeability \( \mu = \mu' - i\mu'' \), and \( H = B/\mu \) is the magnetic field amplitude in the material. The integral is carried out over the volume V of the material.

In this experiment, when the oscillator rotates in its eigenmode, it generates an oscillating field in the external magnetic shield, leading to hysteresis losses. For a magnetic oscillator undergoing small-angle oscillations \( \theta(t) = \theta_0 \sin(\omega_0 t) \), the oscillator can be approximated as a current-carrying coil with an equivalent magnetic moment \( \mathbf{\mu} = i(t) \cdot S = M V \theta(t) \), where \( i \) and \( S \) represent the current and area of the coil, respectively. The dissipation caused by hysteresis effects in the ferrite magnetic shield is then given by 
\begin{equation}
    \gamma_\text{hyst}^\text{shield}=\frac{\bar{P}}{I \langle \dot{\theta}(t)^2 \rangle}=\frac{2\bar{P}M^2V^2}{I \omega_0^2i_0^2S^2}, 
\end{equation}
where $i_\text{0}$ is the amplitude of the current applied to the coil. Finite element simulations reveal that the hysteretic dissipation in the ferrite leads to an oscillator damping coefficient $\gamma_\text{hyst}^\text{shield}/2\pi \approx 0.03~\mathrm{mHz}$.

Additionally, influenced by the large magnet, the small magnet experiences internal hysteresis dissipation during its motion. In this case, the oscillating field acting on the oscillator is given by $B = B_0\theta(t)$, where $B_{0}$ represents the average magnetic field generated by the lifting magnet in the oscillator. For the fully saturated hard ferromagnet oscillator whose $\mu$ is approximately equal to the vacuum permeability $\mu_0$, the intrinsic power loss is $\overline{P}_{\text{hyst}}^{\rm fm}=\frac{1}{2} \omega_0 {\mu}''H^2 V\cite{Vinante2020}=\frac{1}{2 \mu_0^2}\omega_0{\mu}''B_0^2 \theta_0^2 V$, the corresponding hysteresis damping coefficient is 
\begin{equation}
    \gamma_{\rm hyst}^{\rm fm} = \frac{{\mu}'' B_0^2 V}{I \omega_0 \mu_0^2}.
\end{equation}
Since the librational mode resonant frequency of the levitated ferromagnet is given by $\omega_{\text{0}} = \sqrt{\frac{MB_0V}{I}}$, we get
\begin{equation}
    \gamma_{\rm hyst}^{\rm fm} = \frac{{\mu}''I\omega_0^3}{\mu_0^2M^2V}.
\end{equation}

Finite element simulations reveal that with an imaginary magnetic susceptibility (${\mu}''/\mu_0$) of $\sim\!10^{-3}$, magnetic hysteresis dissipation is $\gamma_{\text{hyst}}^{\text{fm}}/2\pi \sim 0.1~\text{mHz}$, exhibiting quantitative agreement with the low-pressure experimental measurements.

\end{document}